\documentclass[aps,pra,twocolumn,floatfix,showpacs]{revtex4-1}
\usepackage{graphicx,amsfonts}
\usepackage{amsmath,amscd,amsfonts,amssymb,color}
\newcommand{\bra}[1]{\left\langle{#1}\right\vert}
\newcommand{\ket}[1]{\left\vert{#1}\right\rangle}

\begin{document}

\title{Teleporting Grin of a Quantum Chesire Cat without cat}

\author{Debmalya Das\(^{1}\) and Arun Kumar Pati\(^{1}\)}

\affiliation{\(^1\)Quantum Information and Computation Group, 
Harish-Chandra Research Institute, HBNI, Chhatnag Road, Jhunsi, Allahabad 211 019, India}

%\lineno

\begin{abstract}
Quantum Chesire Cat is a counterintuitive phenomenon that provides a new window into the nature of the 
quantum systems in relation to multiple degrees of freedom associated with a single physical entity.
Under suitable pre and postselections, a photon (the cat) can be decoupled from its circular polarization
(its grin). In this paper, we explore whether the grin without the cat can be teleported 
to a distant location. This will be a totally disembodied teleportation protocol.
Based on the original Quantum 
Chesire Cat setup, we design a protocol where the circular polarization is successfully teleported between two 
spatially separated parties even though the photon is not physically present with them. The process 
raises questions in our understanding about properties of quantum system. In particular it 
shows that question like ``whose polarization is it'' can prove to be vacuous in such scenario.
\end{abstract}

\maketitle

\section{Introduction}
In the standard measurement scenario of quantum mechanics, the state of a system is disturbed irreversibly
during the process of measurement. 
%When a measurement of an observable $A$ is carried out on a system in the state
%$\ket{\Psi}$, it is projected onto an eigenstate $\ket{a_i}$ of $A$ with a probability given by 
%$|\bra{a_i}\Psi\rangle|^2$. Thus, it is not possible to gain information about the observable $A$, a property of the 
%system, as measurement causes the initial system state $\ket{\Psi}$ to go to one of the eigenstates $\ket{a_i}$.
The outcome of the measurement of an observable is indeterministic and the original state collapses into one of the
eigenstates of the measured observable. This allows us to gain information about the expectation value of the
observable by repeating the measurement over an ensemble of the states.
Weak measurement, on the other hand seeks to gain limited information from the 
quantum system while causing minimal perturbation to the same. As opposed to a projective measurement, this
kind of measurement can be achieved by effecting a weak coupling between the system and the measurement device.

In 1988, Aharonov et. al. proposed the concept of the weak value~\cite{AAV1988},
which is claimed to be the value of an observable $A$, for an ensemble which is initially prepared
in the state $\ket{\Psi_i}$ and finally postselected in the state $\ket{\Psi_f}$. The weak value of a observable
$A$ is obtained in the following way. The system, in the initial state $\ket{\Psi_i}$, is weakly coupled to a suitable
measurement apparatus or meter, thus causing the weak measurement of the observable $A$. This is then followed by the projective
measurement of a second observable $B$ that is incompatible with $A$. In the final step, one of the eigenstates,
$\ket{\Psi_f}$, of the measured observable $B$, is postselected. For all successful postselections of the state 
$\ket{\Psi_f}$, the meter readings corresponding to the weak measurements of $A$ are taken into consideration while
the others are discarded. The shift in the meter readings, on an average, for all such postselected systems is $A_w$ which is
known as the weak value of $A$ and given by
\begin{equation}
 A_w=\frac{\bra{\Psi_f}A\ket{\Psi_i}}{\bra{\Psi_f}\Psi_i\rangle}.
\end{equation}
Clearly, this is a strange value of the observable $A$
that the system reveals between the pre-selection $\ket{\Psi_i}$ and the
post-selection $\ket{\Psi_f}$. In other words, it can be viewed as a property of the system, which
the projective
measurement fails to capture. Some of the fascinating aspects of the weak value are that it can have anomalous 
values that lie outside the eigenvalue spectrum~\cite{AAV1988, Duck1989} and can even be complex
~\cite{Joz2007}.

Although the weak value has been measured experimentally in several quantum systems~\cite{Ritchie1991, Pryde2005,
Lundeen2011, Wu2018, Pal2018}, its meaning has 
ever been a subject of numerous discussions and controversies~\cite{Duck1989, Aharonov1990, Joz2007, Sokolovski2013,
Cormann2016}.
It has been used effectively in signal
amplification and in providing explanations for the spin Hall effect, the three-box paradox and Hardy's 
paradox~\cite{Hosten2008, Tollaksen2010, Dolev2005}. It has also been employed to measure the wavefunction
of a single photon~\cite{Lundeen2011} and to measure the expectation value of non-Hermitian operators~\cite{Pati2015,
Nirala2019}.
Quantum Chesire Cat is a theoretical scheme developed
in~Ref.\cite{Chesire} to challenge some of the 
pre-conceived notions about the nature of a quantum system. It is based on interferometry and asks whether
an intrinsic property, attributed to a system can exist in isolation to the system itself.

In the analysis given in~Ref.\cite{Chesire}, the two properties in question are the position of the photon and its 
polarization. Going by the experience in the classical world, it would seem that a property, like the 
polarization of the photon, can only exist in a region where the photon actually passes through. In other
words, the polarization cannot have an existence independent of the photon itself. The counterintuitiveness
of Quantum Chesire Cat lies in the fact that the photon is detected in one region of the interferometer
while its polarization is detected in a mutually exclusive region. In the next section, we recapitulate 
this phenomenon more rigorously.

The topic of Quantum Chesire Cat has recently drawn a great deal of attention from a large number of researchers
working in quantum information and foundations and has led to a great number of debate and discussions
~\cite{Bancal2013, Duprey2018, Correa2015, Atherton2015}.
In~Ref.\cite{CC} it was argued that the original formulation of 
~Ref.\cite{Chesire} is an incomplete one as it decouples only one component of the polarization from the position.
The former comes up with an alternative interferometric setup that seeks to decouple all the  components 
of polarization degree of freedom from the path degree of freedom. It was also shown in~Ref.\cite{Twin} that more
than two degrees of freedom can be separated in a similar way, a phenomenon called Twin Chesire Cats. The 
three-box paradox, in which a single particle appears with certainty in two disjoint locations, under the 
context of postselection has also been analyzed using Quantum Chesire Cat~\cite{Pan2013}. Recently, in
~Ref.\cite{Richter2018}, the effect has been studied in the presence of decoherence. The phenomenon
has also been observed experimentally using perfect silicon crystal interferometer that separates neutrons from their magnetic 
moments under suitable pre and postselection~\cite{Denkmayr2014}. Other experimental realization of the 
Quantum Chesire Cat can be found in~Ref.\cite{Sponar2016}.

In this paper we ask ` Can we teleport a property without an object?'. To answer the question, we consider the
possibility of using the grin of the Quantum Chesire Cat for teleportation without the cat. 
Using a photon interferometer, we isolate the circular polarization of the photon. We send it to a party, who
teleports it to another spatially separated party using a shared entangled state, local 
operation and classical communication(LOCC). We demonstrate
that although the first party has no photon and no knowledge of the input polarization state, a protocol can 
be designed where it can be successfully teleported to a distant location. At various points, the state 
of the photon polarization can be checked using weak measurements. This can confirm the successful teleportation
of the grin. Using realizations gained from the thought experiment we discuss the implications it has regarding
associating a property of an entity with itself, in the presence of multiple quantum systems.

The paper is organized as follows.
Section II is a recapitulation of the Quantum Chesire Cat effect. In Section III, we present our protocol for 
teleportation of the grin of the Chesire Cat. Section IV deals with some new implications the protocol has 
towards understanding the nature of physical property of the quantum system. We finally conclude with some 
discussions in Section V.

\section{The Quantum Chesire Cat}

\begin{figure}
 \includegraphics{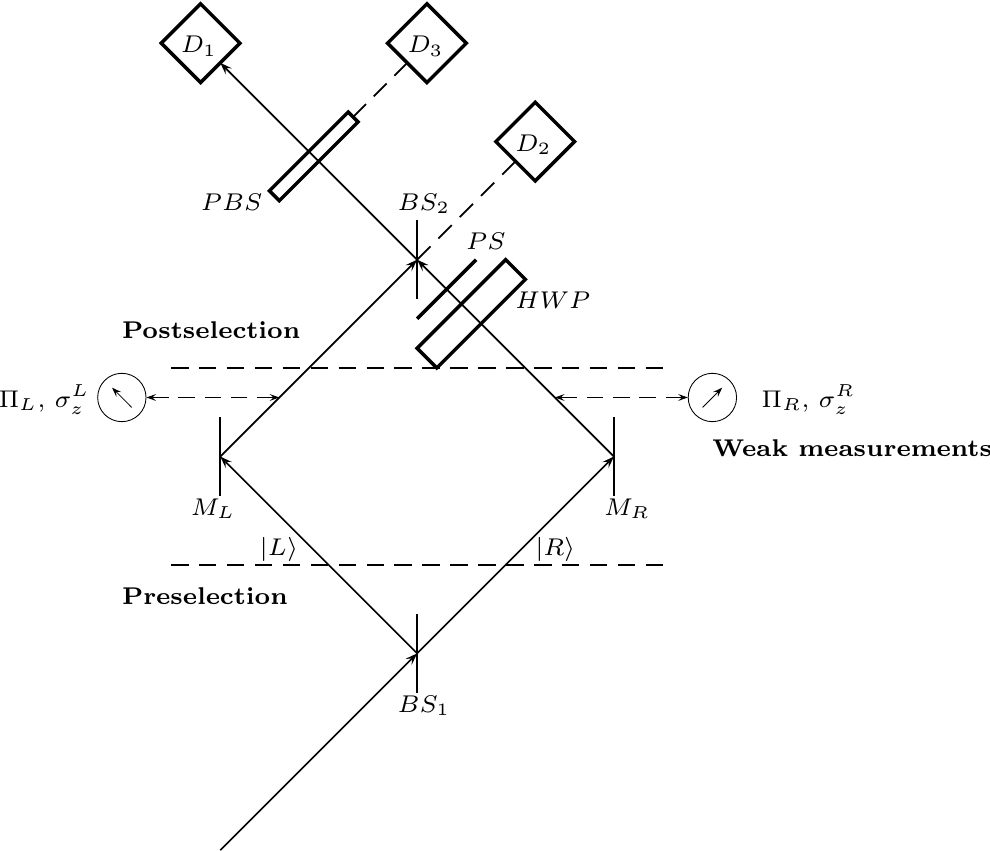}
 \caption{The basic Quantum Chesire Cat setup. The initial state $\ket{\Psi}$ is prepared by passing a 
 photon with linear polarization $\ket{H}$ through a beam-splitter $BS_1$. Weak measurements of positions
 and photon polarizations are carried out on the two arms of the interferometer. The postselection block 
 consists of a half-wave plate $HWP$, a phase-shifter $PS$, the beam-splitter $BS_2$, a polarization beam-
 splitter $PBS$ that transmits polarization states $\ket{H}$ and reflects state $\ket{V}$ and three detectors
 $D_1$, $D_2$ and $D_3$.}
\end{figure}

The phenomenon of Quantum Chesire Cat can be realized by a scheme that is based on a Mach Zehnder interferometer,
first presented in~Ref.\cite{Chesire}. A source sends a linearly polarized single photon towards a 50:50 beam-splitter 
$BS_1$ that channels the photon into a left and right path. Let $\ket{L}$ and $\ket{R}$ denote two orthogonal
states representing the two possible paths taken by the photon, the left and the right arm, respectively. 
If the photon is initially in the horizontal polarization state $\ket{H}$, the photon after passing through the
beam-splitter $BS_1$ can be prepared in the state 
\begin{equation}
 \ket{\Psi}= \frac{1}{\sqrt{2}}(i\ket{L}+\ket{R})\ket{H},
\end{equation}
where the relative phase factor $i$ is picked up by the photon traveling through the left arm  due to the
reflection by the beam splitter. The postselection block, conducting the process of projective measurement and 
eventual postselection, comprises 
of  a half-wave plate (HWP), a phase-shifter (PS), beam-splitter $BS_2$, a polarization beam-splitter (PBS)
 and three detectors $D_1$, $D_2$ and $D_3$. Let the postselected state be
 \begin{equation}
 \ket{\Psi_f}=\frac{1}{\sqrt{2}}(\ket{L}\ket{H}+\ket{R}\ket{V}),
\end{equation}
where $\ket{V}$ refers to the vertical polarization state orthogonal to the initial polarization state 
$\ket{H}$.
The HWP flips the polarization of the photon from $\ket{H}$ to $\ket{V}$ and vice-versa.
The phase-shifter (PS) adds a phase factor of $i$
to the beam. The beam-splitter $BS_2$ is such that when a photon in the state $\frac{1}{\sqrt{2}}(\ket{L}+i\ket{R})$
is incident on it, the detector $D_2$ never clicks. In other words, in such cases, the photon always emerges
towards the PBS. The PBS is chosen such that it always transmits the horizontal polarization
$\ket{H}$ and always reflects the vertical polarization $\ket{V}$. The above arrangement thus ensures that 
only a state that is given by $\ket{\Psi_f}$, before it enters the postselection block, corresponds to the
click of detector $D_1$. Any clicking of the detectors $D_2$ or $D_3$ implies a different state entering the
postselection block. Therefore, selecting the clicks of the detector $D_1$ alone and discarding
all the others leads to the postselection onto the state $\ket{\Psi_f}$.

Define a circular polarization basis as
\begin{eqnarray}
 \ket{+}&=&\frac{1}{\sqrt{2}}(\ket{H}+ i \ket{V}),\nonumber\\
 \ket{-}&=&\frac{1}{\sqrt{2}}(\ket{H}- i \ket{V}).
\end{eqnarray}
Also consider the operator
\begin{equation}
 \sigma_z=\ket{+}\bra{+}-\ket{-}\bra{-}.
\end{equation}
Suppose we want to know which arm a photon, prepared in the state $\ket{\Psi}$ and was ultimately 
postselected in the state $\ket{\Psi_f}$, passed through. This can be effected by performing weak
measurements of the observables $\Pi_L=\ket{L}\bra{L}$ and $\Pi_R=\ket{R}\bra{R}$ by placing weak
detectors in the two arms. Similarly, the polarizations can be detected in the left and the right arms
by respectively performing weak measurements of the following operators
\begin{eqnarray}
 \sigma^L_z=\Pi_L \otimes\sigma_z,\nonumber\\
 \sigma^R_z=\Pi_R \otimes\sigma_z.
\end{eqnarray}
The weak values of the photon positions are measured to be
\begin{equation}
 (\Pi_L)^w=1 \;\mbox{and}\; (\Pi_R)^w=0
 \label{position_wv}
\end{equation}
which implies that the photon in question has traveled through the left arm. The measured weak 
values of the polarization positions, on the other hand, turn out to be
\begin{equation}
 (\sigma^L_z)^w=0\;\mbox{and}\;(\sigma^R_z)^w=1.
 \label{polarization_wv}
\end{equation}
Equations~(\ref{position_wv}) and~(\ref{polarization_wv}) together reveal that the photon traveled 
through the left arm but its circular polarization traveled through the right arm. This means
the two degrees of freedom of a single entity can, in fact, be decoupled. That is, a property of a quantum
system can exist independent of its existence in that region.

%\section{Partial Quantum Chesire Cat}

%It is interesting to study a variant of the above formulation by choosing a different postselected state.
%Suppose we carry out the postselection of the following state
%\begin{equation}
% \ket{\Psi_f^P}=\frac{1}{\sqrt{2}}(\ket{L}+\ket{R})(\gamma\ket{H}+\delta\ket{V},
%\end{equation}
%where $\gamma$ and $\delta$ are real numbers.

%Performing the weak measurements of $\Pi_L$, $\Pi_R$, $\sigma_z^L$ and $\sigma_z^R$ we obtain

%\begin{eqnarray}
% (\Pi_L)^w &=& \frac{1}{2}+\frac{i}{2},\nonumber\\
%  (\Pi_R)^w &=& \frac{1}{2}-\frac{i}{2},\nonumber\\
%  (\sigma_z^L)^w &=& 0,\nonumber\\
%  (\sigma_z^R)^w &=& \frac{1}{2}\frac{\delta}{\gamma}+\frac{i}{2}\frac{\delta}{\gamma}.
%\end{eqnarray}

%This means that the photon has travelled through both the arms of the interferometer but
%surprisingly the polarization has travelled only throught the right arm. We call this 
%partial Quantum Chesire Cat.

%Thus Quantum Chesire Cat is one of the many counterintuitive phenomena observed in the 
%setup although it is arguably the most visible and interesting as it exhibits a complete
%separation of the photon from its polarization. By trying various combinations of the 
%preselected and the postselected states, one can obtain several versions of the partial
%Quantum Chesire Cat.

\section{Teleportation using the decoupled polarization}

%It is interesting to investigate whether one can perform quantum information protocols with a 
%died degree of freedom. As an example, we try to perform teleportation with a photon
%polarization
In quantum teleportation, one can recreate the quantum state at a distant location using entanglement,
local operation and classical communication. However, the particle is present at one end of the shared entangled state,
with the teleporter, where the Bell-state measurement is carried out. \textit{Whereas here we will discuss the 
teleportation of 
the photon polarization, while the photon itself is at a different place}. Consider four parties Alice, 
Bob, Charlie and Dave, who are all spatially separated from each other. The setup is primarily 
based on the already discussed Mach-Zehnder interferometer arrangement of the Quantum Chesire
Cat. Charlie prepares the initial state and Dave performs the postselection. Alice and Bob are 
situated on an arm of the interferometer through which the disembodied polarization state travels, 
conditioned to the appropriate postselection by Dave. We would like to test whether it is possible for Alice to 
teleport the polarization of a photon to Bob, while the photon is not physically present with her.
The initial linear  polarization is taken to be in an arbitrary direction
\begin{equation}
 \ket{\psi}=\alpha\ket{H}+\beta\ket{V}.
\end{equation}.

\begin{figure}
 \includegraphics{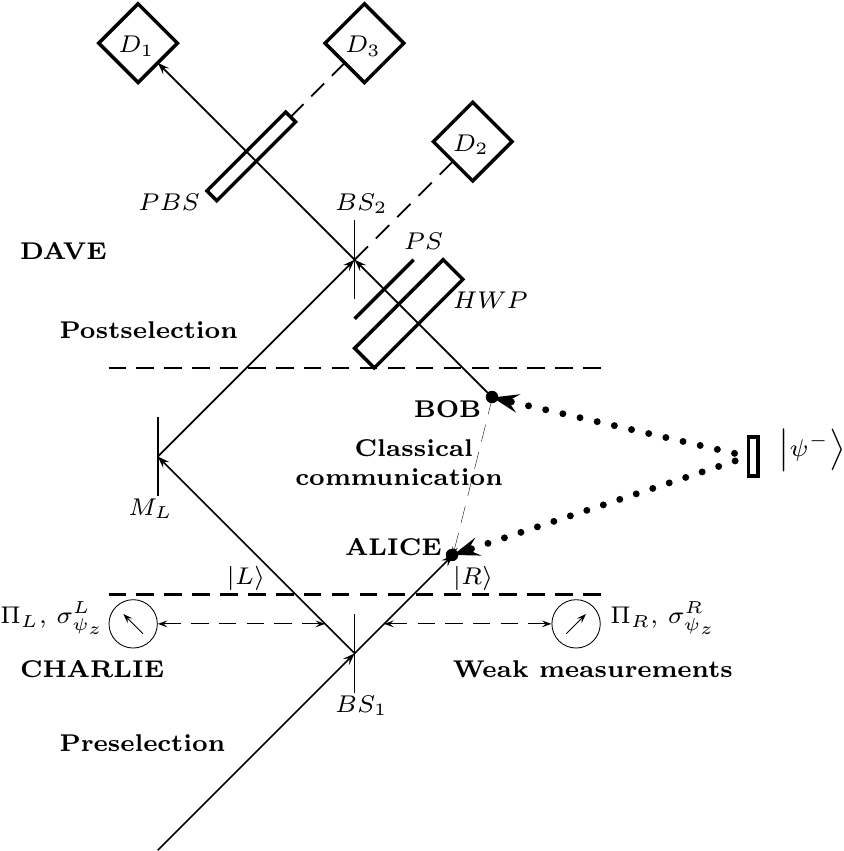}
 \caption{Quantum teleportation of the grin of a Quantum Chesire Cat. The process involves four parties
 Charlie, Alice, Bob and Dave. Charlie prepares the initial state $\ket{\Psi}$ while Dave performs the 
 postselection in the state $\ket{\Psi_f}$. The isolated grin or the photon polarization state $\ket{\psi}$
 is sent to Alice, located on the right arm of the interferometer. Alice shares with Bob, who is at a distant 
 location but on the same arm, a singlet state $\ket{\psi^-}$. By performing local operations jointly on the 
 polarization state and the shared state, Alice classically communicates her outcomes to Bob, who then 
 reproduces the initial polarization by applying local unitaries. Weak measurements can be performed by Charlie
 or by Dave to check the isolated state.}
\end{figure}

Let the initial state $\ket{\Psi}$, prepared by Charlie, using $BS_1$ be
\begin{equation}
 \ket{\Psi}= \frac{1}{\sqrt{2}}(i\ket{L}+\ket{R})\ket{\psi}.
\end{equation}
He sends the photon and its 
polarization to Dave who is in possession of the postselection block, consisting of the HWP,
PS, $BS_2$, PBS and the detectors $D_1$, $D_2$ and $D_3$. It is Dave who postselects the state 
$\ket{\Psi_f}$, given by
\begin{equation}
 \ket{\Psi_f}=\frac{1}{\sqrt{2}}(\ket{L}\ket{\psi}+\ket{R}\ket{\psi^\perp}),
\end{equation}
where $\ket{\psi^\perp})$ represents the polarization state, orthogonal to the initial linear
polarization $\ket{\psi}$. Thus,
\begin{equation}
 \ket{\psi^\perp}=\beta^*\ket{H}-\alpha^*\ket{V}.
\end{equation}
This prompts us to define a new basis for circular polarization,
\begin{eqnarray}
\ket{\psi_+}=\frac{1}{\sqrt{2}}(\ket{\psi}+i\ket{\psi^\perp}),\nonumber\\
\ket{\psi_-}=\frac{1}{\sqrt{2}}(\ket{\psi}-i\ket{\psi^\perp}),
\end{eqnarray}
with $\ket{\psi_+}$ and $\ket{\psi_-}$ being the eigenstates of the operator
\begin{equation}
 \sigma_{\psi_z}=\ket{\psi_+}\bra{\psi_+}-\ket{\psi_-}\bra{\psi_-}.
\end{equation}
Clearly, $\sigma_{\psi_z}$ is related to $\sigma_z$ by
\begin{equation}
 \sigma_{\psi_z}=U^\dagger \sigma_z U,
\end{equation}
where $U$ is a unitary operator. The measurement of the polarization state of the photon
is carried out by measuring the operator $\sigma_{\psi_z}$. More specifically, the weak 
measurements of $\sigma^L_{\psi_z}$ and $\sigma^R_{\psi_z}$ are carried out in the left and right arms of
the interferometer
by Charlie to find out which way the polarization went without disturbing the state.
Here, $\sigma^L_{\psi_z}$ and $\sigma^R_{\psi_z}$ are, respectively, defined as
\begin{eqnarray}
 \sigma^L_{\psi_z}=\Pi_L \otimes\sigma_{\psi_z},\nonumber\\
 \sigma^R_{\psi_z}=\Pi_R \otimes\sigma_{\psi_z}.
\end{eqnarray}
It is noteworthy that Charlie, and not Alice or Bob, conducts this measurement of photon
polarization, since it is impossible for Alice or Bob to know the initial polarization state
unless this information is shared by Charlie. Hence, Alice and Bob are unaware of the basis 
required for defining the operator $\sigma_{\psi_z}$. They can always gain information about 
the polarization state using projective measurement, but that will amount to disturbing the 
system and jeopardizing the whole process of teleportation of the disembodied polarization.

Now, consider the Bell states in the $\{\ket{H}, \ket{V}\}$ basis as, given by
\begin{eqnarray}
 \ket{\phi^+}=\frac{1}{\sqrt{2}}(\ket{HH}+\ket{VV}),\nonumber\\
 \ket{\phi^-}=\frac{1}{\sqrt{2}}(\ket{HH}-\ket{VV}),\nonumber\\
 \ket{\psi^+}=\frac{1}{\sqrt{2}}(\ket{HV}+\ket{VH}),\nonumber\\
 \ket{\psi^-}=\frac{1}{\sqrt{2}}(\ket{HV}-\ket{VH}).
 \label{Bell_basis}
\end{eqnarray}

From the previous discussion we know that, for the postselected states, the photon
travels through the left arm while the circular polarization 
goes via the right arm. As discussed earlier, the two parties Alice and Bob, who are spatially separated,
occupy two positions in the right arm. Alice receives the polarization state $\ket{\psi}$, unknown to her,
and is required to communicate it to Bob. This can be achieved in the following way. Alice and Bob
mutually share a singlet polarization state
$\ket{\psi^-}=\frac{1}{\sqrt{2}}(\ket{HV}-\ket{VH})$
between each other. This state is appropriate for teleportation, as it remains invariant up to a phase
with the change is basis. This is important because for Alice and Bob the pure state $\ket{\psi}$ they
would like to teleport is unknown. Thus the basis defining the operator $\sigma_{\psi_z}$ is also
unknown to them. Consequently, Alice and Bob are free to define their mutually shared state $\ket{\psi^-}$
in the $\{\ket{H}, \ket{V}\}$ basis due to their ignorance of the state of the polarization
to be teleported.

On the joint state consisting of the polarization state, sent by Charlie, and 
the shared singlet state, Alice performs a local joint Bell measurement and projects
her polarization state into one of the states given by Equation~(\ref{Bell_basis}).
For example, if the outcome of Alice is $\ket{\psi^+}$ then the total state consisting
of the input state and the joint state she shares with Bob now becomes 
$\ket{\psi^+}(-U_z\ket{\psi})$ with
\begin{equation}
 U_z=\ket{H}\bra{H}-\ket{V}\bra{V}.
 \label{random}
\end{equation}
Alice then classically communicates her outcome to Bob.
To reproduce the state $\ket{\psi}$, at his end, Bob applies
the $U_z$ operator locally and achieves the same state $\psi$ upto a minus sign. 
Similarly, depending upon Alice's outcomes being $\ket{\phi^+}$, $\ket{\phi^-}$ or $\ket{\psi^-}$, 
which are classically communicated to Bob, he subsequently applies
one of the local operators $U_y=-i \ket{H}\bra{V} + i \ket{V }\bra{H}$, $U_x=\ket{H}\bra{V}+\ket{V}\bra{H}$ or $\mathbb{I}$
on the state he shares with Alice. In doing so the polarization state $\ket{\psi}$ is
reproduced at Bob's end and is then sent to Dave who proceeds with the strong measurement and 
postselection as described before.

Charlie can install two detectors, one weakly measuring $\Pi_L$ and the other weakly measuring
$\sigma_{\psi_z}^R$ on the left and right arm of the interferometer, respectively, to check whether 
the photon is traveling through the left arm and the circular polarization is traveling through the right arm,
as before, using the corresponding weak values, for all successful postselections of $\ket{\Psi_f}$.
The results are same as obtained earlier. Thus,
\begin{equation}
 (\Pi_L)^w=1 \;\mbox{and}\; (\Pi_R)^w=0
 \label{position_arb_wv}
\end{equation}
Also,
\begin{equation}
 (\sigma^L_{\psi_z})^w=0\;\mbox{and}\;(\sigma^R_{\psi_z})^w=1.
 \label{polarization_arb_wv}
\end{equation}
which means that Charlie has checked that the circular polarization has been sent through
the right arm, and the photon through the left arm, for all cases in which the state $\ket{\Psi_f}$ is
postselected in future.
This circular polarization is teleported by Alice to Bob, while the photon itself is in the left arm.
Dave can also install weak detectors and double check whether Bob actually recreated the input polarization.
It is, however, imperative that Alice and Bob do not perform any measurement, strong or weak, to gain
knowledge about the state or to check the success of teleportation. This is because, strong measurement will
rupture the whole process by disturbing the system and weak measurements need to be performed in a specific 
basis, unknown to both Alice and Bob.

\section{Whose grin makes it to the end?}

We have shown that the circular polarization of a photon can be teleported while the photon itself is
at some other location. In this section we throw light on another curious aspect pertaining to the process.
Notice that when the polarization arrives at Alice's port, true to the spirit of teleportation,
she does not physically transport the polarization to Bob's port. Instead, the shared EPR state between 
Alice and Bob is converted to the input state at Bob's end by virtue of the operations done by Alice and 
Bob. It is this polarization state, that Bob sends to Dave for combination with the spatial degree of 
freedom, traveling through the other arm. This begs us to ask the question, if the grin of the cat 
ends its journey at Alice's port, whose grin recombines with the cat at Dave's location? It would seem
that this grin belongs to one of the polarization degrees of freedom of the EPR state, Bob's subsystem of 
the polarization singlet state in our case. In other words we have exchanged the grin of two Quantum Chesire 
Cats.

There are more counterintuitive aspects to this process. The grin or the polarization derived from the original
input state ends its journey with Alice. Alice thus has at her disposal the polarization from the input state
and the spatial and polarization degrees of freedom from the shared EPR state. On the other hand, Bob is
left with the spatial degree of freedom obtained from the EPR state. Thus, at the end of the entire process,
Alice has a polarization without the photon while Bob has a photon without a polarization. This situation
is a more permanent decoupling of the cat and its grin as opposed to the usual Quantum Chesire Cat scenario
in which the grin recombies with the cat at Dave's port.

It must be remembered that all the anomalous effects discussed so far, starting from the separation of 
the photon and its polarization, the teleportation of the polarization alone to the permanent decoupling
of the photon and the polarization, are all in the context of a successful postselection. For all other 
outcomes the process proceeds as expected without the separation of the cat and its grin.

\section{Conclusions}

%We have used 
To summarize, using the original Quantum Chesire Cat setup 
%and isolated the photon polarization  from the photon itself.
we have 
%We have then made modifications to the setup and
used the isolated photon polarization to perform teleportation even when the photon is not present.
It has been shown that our protocol does not require a knowledge of the polarization state for the teleporter.
But it is also revealed that the parties participating in the teleportation must remain ignorant of the original
state of the photon polarization if the information if not shared with them a priori.
We have also hinted at the counterintuitiveness of different photons exchanging their polarizations.
%Additionally we have also demonstrated that presence of noise in the teleportation channel can result in
%a partial photon carrying the entire polarization through the teleportation channel.

With the success in performing quantum teleportation with the grin of the Quantum Chesire Cat, it would be 
interesting to explore the possibilities of performing other quantum information processing tasks with the 
same. We can speculate that the separation of a quantum system from its intrinsic property may lead to 
a greater security in future quantum communications but it requires further exploration.
We are also currently expanding upon the idea of swapping of grins of two Quantum Chesire Cats. Thus, in the 
quantum world, property of a quantum system cannot be claimed to be its own. It could be someone else's property
that a particle owns momentarily.

\end{document}